\documentclass[12pt]{iopart}

\usepackage{iopams}
\usepackage{color}
\usepackage{graphicx}

\begin{document}

\title[]{Phase coherent transport and spin-orbit interaction in GaAs/InSb core/shell nanowires}

\author{Patrick Zellekens$^{1,4,5}$, Natalia Demarina$^{2,4}$, Johanna Jan\ss en$^{1,4,5}$, Torsten Rieger$^{1,4}$, Mihail Ion Lepsa$^{1,4}$,  Pujitha Perla$^{1,4,5}$, Gregory Panaitov$^{3,4}$  Hans L\"uth$^{1,4}$, Detlev Gr\"utzmacher$^{1,4}$, Thomas Sch\"apers$^{1,4}$}

\address{$^1$ Peter Gr\"unberg Institute (PGI-9), Forschungszentrum J\"ulich, 52425 J\"ulich, Germany}

\address{$^2$ Peter Gr\"unberg Institute (PGI-2), Forschungszentrum J\"ulich, 52425 J\"ulich, Germany}

\address{$^3$ Institute of Complex Systems (ICS-8), Forschungszentrum J\"ulich, 52425 J\"ulich, Germany}

\address{$^4$ JARA-Fundamentals of Future Information Technology, J\"ulich-Aachen Research Alliance}

\address{$^5$ Experimental Physics IV F, RWTH Aachen University, 52056 Aachen, Germany}

\ead{th.schaepers@fz-juelich.de}

\hyphenation{mag-ne-to-trans-port nano-dots mag-ne-to-con-duc-tance}

\begin{abstract}
Low-temperature magnetotransport measurements are performed on GaAs/InSb core-shell nanowires. The nanowires were self-catalyzed grown by molecular beam epitaxy. The conductance measurements as a function of back-gate voltage show an ambipolar behavior comprising an insulating range in between the transition from the $p$-type to the $n$-type region. Simulations based on a self-consistent  Schr\"odinger--Poisson solver revealed that the ambipolar characteristics originate from a Fermi level dependent occupation of hole and electron states within the approximately circular quantum well formed in the InSb shell. By applying a perpendicular magnetic field with respect to the nanowire axis, conductance fluctuations were observed, which are used to extract the phase-coherence length. By averaging the magneto-conductance traces at different back-gate voltages, weak antilocalization features are resolved.  Regular flux-periodic conductance oscillations are measured when an axial magnetic field is applied. These oscillations are attributed to closed-loop quantized states located in the InSb shell which shift their energetic position periodically with the magnetic flux. Possible reasons for experimentally observed variations in the oscillation patterns are discussed using simulation results.     
\end{abstract}

\maketitle

\section{Introduction}\label{sec:introduction} 

Nanostructures based on InSb, in particular nanowires, have attracted considerable interest because of the small electron effective mass and the related very high electron mobility of InSb. This makes this material very interesting for high-speed nanoelectronic applications \cite{Candebat09,Nilsson10,Wang11,Thelander12,Guel15}. Furthermore, the very small band gap, i.e. about 180\,meV at 300\,K, results in an ambipolar behaviour by changing the gate voltage in a field-effect transistor allowing for switching between $p$- and $n$-type transport \cite{Candebat09,Rieger17}. 
Because of the very large g-factor and the presence of spin-orbit coupling, spin-related coherent quantum transport can be studied in detail and applications with respect to spintronic devices and topological quantum computation are suggested \cite{Nilsson09,Weperen15,Weperen15,Gazibegovic17}.
Regarding the latter, evidence of the presence of Majorana zero modes was reported in structures based on InSb nanowires partly covered by a superconducting electrode \cite{Mourik12,Zhang17}. However, recent theoretical predictions suggest, that the phase space in which Majorana zero modes could form can be significantly enhanced in terms of the required strong spin-orbit interaction, if a narrow-gap semiconductor like InAs or InSb is combined with a wide-gap semiconductor as a core material \cite{Woods19}. This happens due to the shift of the electronic states towards the outer surface of the core/shell system, at which the slope of the bands and the corresponding electrical field is much larger.

In the present paper, GaAs/InSb core-shell nanowires with a highly conducting narrow-gap InSb shell and a high-resistive wide-gap GaAs core are studied by magneto-conductance measurements at low temperature. The tubular shell topology of the conductive channel offers the advantage of good control of the action of the magnetic field and of the gate voltage on the circulating electron current. From previous studies of the defect structure and transport properties at room temperature, the InSb shell is revealed to be fully relaxed due to the high lattice mismatch between GaAs and InSb \cite{Rieger17}. Thus, for the present theoretical analysis standard bulk values of all interesting properties of the InSb shell can be assumed in good approximation. 

For the future realization of devices based on nanowire/superconductor devices, the underlying core/shell nanowire system has to fulfill a couple of benchmark criteria, including a sufficiently long phase-coherence length $l_\varphi$. In prior magneto-conductance measurements on GaAs/InAs core-shell nanowires, clear Aharonov--Bohm-type oscillation patterns could be resolved when a magnetic field was threading the cross section of the nanowire \cite{Guel14,Haas16,Haas17}. For the present GaAs/InSb core-shell nanowires, the situation is more complex because of the stronger effect of the Zeeman energy due to the much larger $g$-factor of InSb compared to InAs. By performing transport measurements in a parallel magnetic field, we analyzed the resistance modulations with respect to Aharonov--Bohm-type oscillations and a possible influence of Zeeman-splitting on the energy dispersion of the flux-periodic energy spectrum.  In order to gain information on $l_\varphi$, we also performed transport measurements in a magnetic field oriented normal to the wire axis. By analyzing the resulting universal conductance fluctuations (UCFs) spectrum in terms of the correlation field, we obtain values for $l_\varphi$ in wires with different geometries. One of the key differences to our previous nanowire system based on GaAs/InAs is the occurance of weak antilocalization (WAL), indicating the presence of spin-orbit coupling. By using a combination of gate-averaging and a fitting model, we extract values for the spin-orbit scattering length $l_{SO}$.

\section{Experimental}\label{sec:experimental}

The GaAs/InSb core/shell nanowires were grown in two steps using molecular beam epitaxy. Details on the growth parameters can be found in Ref.~\cite{Rieger17}. First, GaAs nanowires were grown from Ga droplets formed in  pinholes of a thin layer of native silicon oxide on a Si (111) substrate. In the subsequent step the InSb shell was grown on the sidewall facets of the GaAs core nanowire. Here, a nucleation of InSb V-shaped islands on the side facets of GaAs nanowires was followed by a coalescence of the islands leading to the formation of an InSb shell. The (average) shell thickness increases linearly with the growth time and the growth rate is about $20 - 25$\,nm/h. We investigated samples of two growth runs, i.e. sample A with a core radius $r_c = 25$\,nm and a shell thickness of $t_\mathrm{s}=50$\,nm (growth run A) and samples B1 and B2 (growth run B) with $r_c = 35$\,nm and $t_\mathrm{s}=22 - 30$\,nm. Due to the shorter growth time, the shell of the second growth run has a rather pronounced roughness leading to the non-uniform average shell thickness. In contrast, the surface of the nanowires of the first growth run (sample A) is smoother and the shell thickness is nearly uniform along the nanowire. The GaAs core nanowires have a hexagonal morphology and mainly consist out of a zinc blende crystal structure, which is adopted by the InSb shell. In a previous study we could show, that the large lattice mismatch between GaAs and InSb of approximately 14\% and the according strain is compensated by forming a high density of misfit dislocations at the interface \cite{Rieger17}. By performing a geometric phase analysis based on high-resolution transmission electron microscopy, it could be shown, that the strain is relaxed abruptly at the interface within a distance of 1.5$\,$nm, i.e. we don't expect any significant effect of strain on the band structure.   
\begin{figure}[htb]
	\begin{center}
		\includegraphics[width=0.9\textwidth]{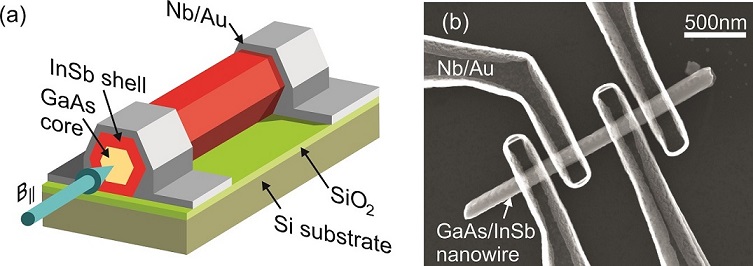}
		\caption{(a) Schematic illustration of the sample layout and the orientation of the parallel magnetic field $B_{||}$. (b) Scanning electron micrograph of sample B1. The nanowire surface shows some roughness.}
		\label{fig:scheme-sample}
	\end{center}
\end{figure}

GaAs/InSb core-shell nanowires were contacted individually by first transferring them to a highly $n$-doped silicon substrate with a 200-nm-thick SiO$_2$ layer. Next, the contact fingers with a width of 250\,nm were defined by electron beam lithography. The contact areas were thereby cleaned by means of Ar$^+$ sputtering prior to the metal deposition. Subsequently, a Nb/Au  (3\,nm/60\,nm) layer was sputter-deposited and lift-off was performed to obtain ohmic contacts. The contacts had a typical separation of 500\,nm. A schematic view of the contacted sample is depicted in figure~\ref{fig:scheme-sample}(a), while figure~\ref{fig:scheme-sample}(b) shows a scanning electron microscopy image of sample B1. The magneto-transport measurements were performed in a He-4 variable temperature insert with a minimum base temperature of 1.4\,K.  Magnetic fields up to 13\,T were available. In some measurements the highly $n$-doped substrate served as a back-gate electrode to modulate the carrier concentration.The transport measurements were performed in a two-terminal configuration by employing a lock-in technique. We used an ac voltage bias of 40\,$\mu$V for the magnetotransport measurements, while the current was measured using a current-voltage converter. The contact resistance was low, i.e. in the range of 20 to 100$\,\Omega$, and was neglected.

\section{Results and Discussion}\label{sec:results} 

\subsection{Ambipolar Behaviour} \label{sec:ambipolar} 

\begin{figure}[htb]
	\begin{center}
	    \includegraphics[width=0.75\textwidth]{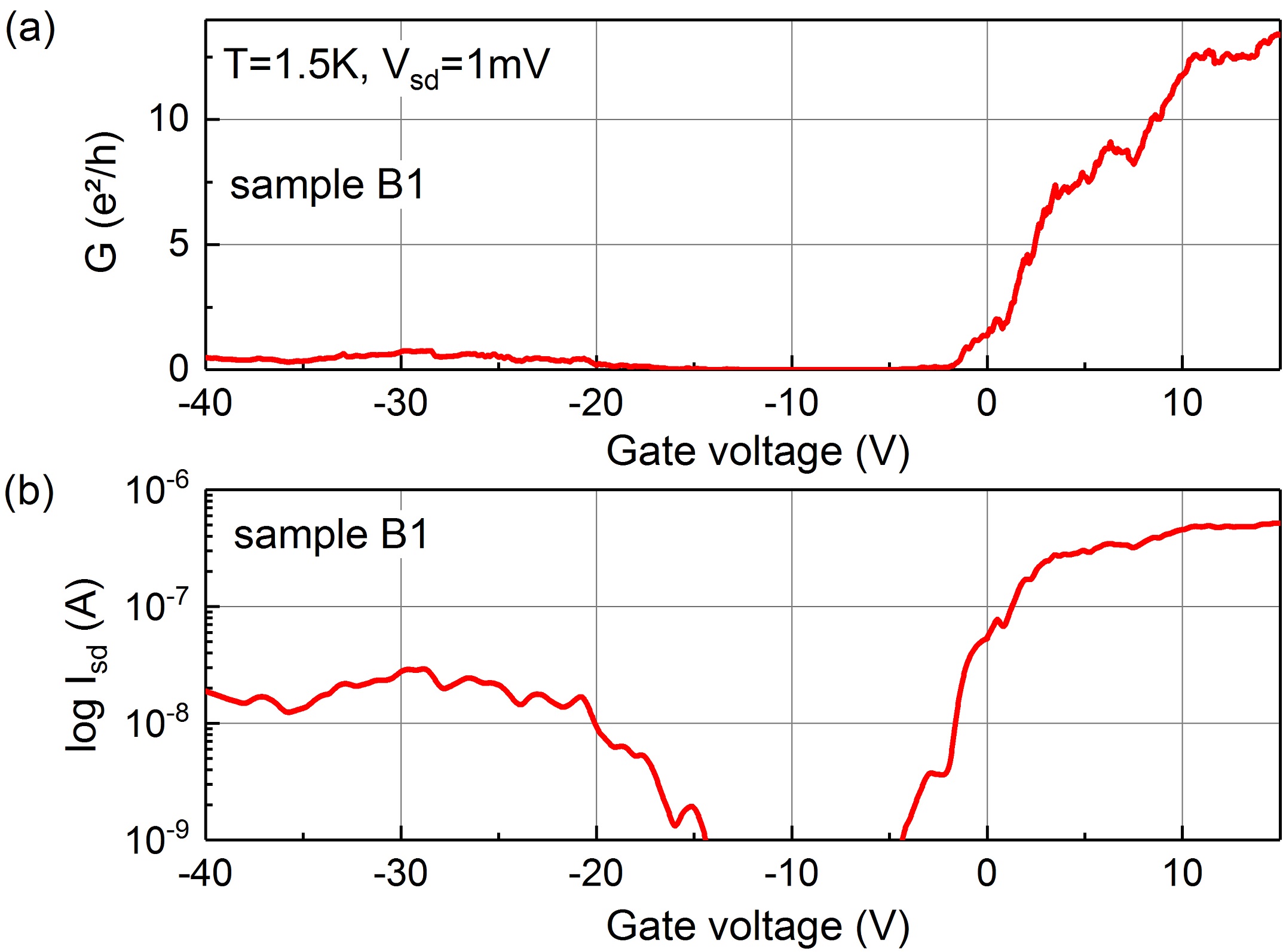}
		\caption{(a) Conductance $G$ of sample B1 in units of $e^2/h$ as a function of back-gate voltage $V_\mathrm{g}$. The measurement was performed at $T=1.5\,$K. (b) Logarithmic plot of the corresponding source-drain current $I_\mathrm{sd}$ vs. $V_\mathrm{g}$. }
		\label{fig:ambipolar}
	\end{center}
\end{figure}

General information on the type of transport in our GaAs/InSb core-shell nanowires is gained by measuring the conductance as a function of back-gate voltage $V_\mathrm{g}$. Figure~\ref{fig:ambipolar}(a) shows an exemplary measurement on sample~B1 at 1.5\,K for a source-drain bias voltage of $V_\mathrm{sd}=1$\,mV, which was applied by means of a DC voltage source. Corresponding data for samples of growth run A are given in the Supplementary Material.  
Two conductive ranges can be identified. First, at a gate voltage of about $-4$\,V, the conductance increases by increasing $V_\mathrm{g}$, thus indicating $n$-type conductance of the nanowire. Second, from a gate voltage of approximately $-15$\,V the conductance increases as well towards more negative gate voltages, which corresponds to a $p$-type behaviour. Between these branches, the conductance is suppressed. We attribute this to an ambipolar behaviour, i.e. a switching between $n$- and $p$-type transport, induced by shifting the Fermi level $E_\mathrm{F}$ from the conduction band into the valence band by decreasing $V_\mathrm{g}$. In between $E_\mathrm{F}$ is passing the band gap, thus no carriers are available for transport. The ambipolar behavior occurs because of the very small band gap of InSb, i.e. 235\,meV at 1.5\,K, compared to other III-V semiconductors. In order to point out the rapid decrease of the conductance when the band gap region is approached, we also plotted the source-drain current in a logarithmic scale in figure~\ref{fig:ambipolar}(b). Such an ambipolar transfer characteristic was observed before for InSb nanowire-based field-effect transistors \cite{Candebat09,Rieger17}. However, since these measurements were performed at room temperature, no complete pinch-off was observed in the band-gap region because of thermal smearing. In contrast, in our case at a temperature of 1.5\,K a full pinch-off is achieved. As can be seen in figure~\ref{fig:ambipolar}(b), the conductance in the $p$-type region is lower than in the $n$-type region. We attribute this to the larger effective hole mass compared to the electron mass and the according lower mobility for holes. Furthermore, we cannot rule out that the contact resistance is increased for the $p$-type region, since the measurements are performed in a 2-terminal configuration. A closer look on the transfer characteristics reveals some superimposed modulations. At small conductance values, at the threshold to the gap region, the current-voltage characteristics revealed a non-linear behaviour with a plateau around zero source-drain bias, which we attribute to Coulomb blockade effects \cite{Weis14}. Owing to the low carrier concentration and potential fluctuations caused by disorder, charge islands separated by barriers are formed. In case of electron transport, a linear current-voltage characteristic was achieved at larger positive gate biases, e.g. at $V_\mathrm{g} > 5$\,V for the sample shown in figure~\ref{fig:ambipolar}. In the $p$-type transport regime we did not succeed in observing a linear characteristics. As a consequence, the following magneto-transport measurement were performed exclusively at a fixed positive gate bias in the $n$-type region, in order to prevent single electron transport contributions.  

\begin{figure}[htb]
	\begin{center}
		\includegraphics[width=0.9\textwidth]{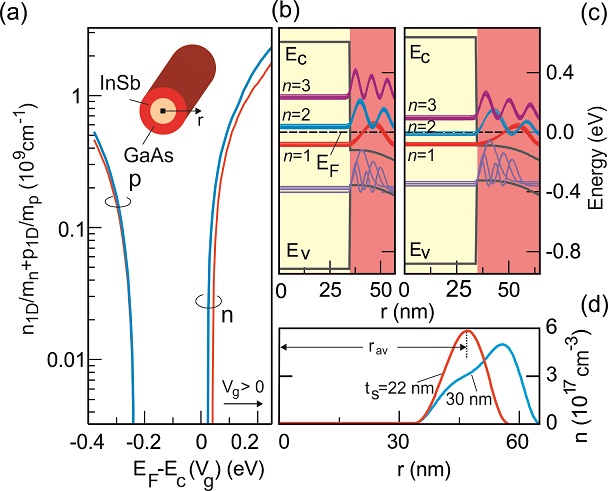}
		\caption{(a) $G_\mathrm{t} \propto n_\mathrm{1D}/m_n + p_\mathrm{1D}/m_p$ as a function of the conduction band edge position with respect to $E_\mathrm{F}$ at the nanowire surface. The simulations correspond to nanowires of growth run B, assuming an InSb shell thickness of 22\,nm (red) and 30\,nm (blue), respectively. The Fermi level is set to zero energy. Inset: Sketch of the GaAs/InSb core-shell nanowire with a cylindrical cross section. (b) and (c) Band diagram and squared eigenfunctions for a GaAs/InSb nanowire corresponding to the second growth run   with the shell thickness set to 22~nm (b) and 30~nm (c) and $E_\mathrm{F}-E_\mathrm{c}=0.2$\,eV;  (d) Three-dimensional electron density as a function of the radial coordinate $r$ for a nanowire corresponding to growth run B with the shell thickness set to 22\,nm (red) and 30\,nm (blue) and $E_\mathrm{F}-E_\mathrm{c}=0.2$\,eV.}
		\label{fig:simulation-density}
	\end{center}
\end{figure}

For the theoretical interpretation of the measurements shown in figure~\ref{fig:ambipolar} we performed band structure calculations. Here, we approximated the hexagonal cross section by a cylindrical one (cf. figure~\ref{fig:simulation-density}(a) (inset)) to simplify the simulations. The coupled Schr\"odinger and Poisson equations were solved self-consistently for the envelope function $\Phi_{nl}(r) \exp (il\varphi)$ within the effective mass approximation and by assuming a lattice temperature of 4\,K, with $r$ and $\varphi$ as the radial and angular coordinates, respectively, and $n$ and $l$ the radial and the orbital quantum numbers, respectively \cite{Demarina13}.
The measured source-drain current as a function of the gate voltage, as plotted in  figure~\ref{fig:ambipolar}, does not show indications of quantized conductance due to ballistic transport. Instead, because of the numerous crystal lattice imperfections, the source-drain current undergoes random fluctuations. Thus, diffusive charge transport can be assumed. As a consequence, we assume for the qualitative analysis that the electrical conductivity is determined by the sum of the product of the electron and hole mobilities, $\mu_n$ and $\mu_p$ and the total electron and hole densities, $n_\mathrm{1D}$ and $p_\mathrm{1D}$, respectively. The latter values were calculated by integrating the three-dimensional electron and hole densities, i.e. $n(r)$ and $p(r)$, over the nanowire cross section. The total conductance is defined as $G_\mathrm{t}=e (n_\mathrm{1D} \mu_n + p_\mathrm{1D} \mu_p)$.  Following the relaxation time approximation and neglecting the origin of the charge carrier scattering, we can assume, that $G_\mathrm{t} \propto n_\mathrm{1D}/m_n + p_\mathrm{1D}/m_p$, where $m_n$ and $m_p$ are the effective mass of electrons and holes, correspondingly. In order to simulate the effect of the back-gate voltage on the nanowire conductance we vary the boundary condition for the electrostatic potential at the nanowire surface, assuming that the electric field of the back-gate voltage does not break the axial symmetry. If the nanowire is subject to an applied gate voltage $V_\mathrm{g}$, the position of the conduction band edge at the nanowire surface reads $E_\mathrm{c}(V_\mathrm{g})=E_\mathrm{c}(0)- \kappa eV_\mathrm{g}$, where $E_\mathrm{c} (0)$ is the unbiased position of the conduction band edge at the nanowire surface and $\kappa$ is the coefficient which determines the ratio between the position of the conduction band with respect to the Fermi level $E_\mathrm{F}$ used for the calculations and the applied gate voltage. The latter will be discussed below.

Figure~\ref{fig:simulation-density}(a) shows $G_\mathrm{t} \propto n_\mathrm{1D}/m_n + p_\mathrm{1D}/m_p$ as a function of conduction band edge position $E_\mathrm{c}(V_\mathrm{g})$ with respect to $E_\mathrm{F}$ at the nanowire surface. The calculations were performed based on the minimum and maximum shell thickness of 22 and 30\,nm, respectively, according to the geometry of nanowires of growth run B. The right part of the curve corresponds to a positive applied gate voltage and an $n$-type conductance of the nanowire. Obviously, the large positive gate voltage causes a downward band bending with $E_\mathrm{F}$ positioned inside the conduction band of the InSb shell. This case is demonstrated in figure~\ref{fig:simulation-density}(b) for $t_\mathrm{s}=22$\,nm. The electron states of the first subband ($n=1$) are below $E_\mathrm{F}$ and populated by electrons, thus electron accumulation at the nanowire surface occurs and the shell has an $n$-type conductivity (cf. figure~\ref{fig:simulation-density}(a)). The electron density distribution shown in figure~\ref{fig:simulation-density}(d) has a maximum near the center of the InSb shell. The intrinsic GaAs core of the nanowire is depleted, i.e. all the transport is restricted to the InSb shell. This assumption is based on the previous findings presented in \cite{Rieger17}, which showed a linear dependency between shell thickness and measured conductance for wires of different growth runs, which was independent from the corresponding core diameter. The change in the gate voltage towards the negative value leads to an upward bending of the energy bands at the surface. At a certain gate voltage the Fermi level is positioned inside the band gap which at low temperature causes a depletion of the nanowire. A further upward band bending causes a Fermi level shift into the valence band leading to a $p$-type conductance of the nanowire (cf. figure~\ref{fig:ambipolar} and figure~\ref{fig:simulation-density}(a)). The $p$-type conductance is smaller than the corresponding $n$-type conductance due to the larger hole effective mass which agrees well with the experimental data shown in figure~\ref{fig:ambipolar}. 

As mentioned above, the nanowires of growth run B have a shell which is non-uniform in width along the nanowire axis varying between 22 and 30\,nm. Thus, in figure~\ref{fig:simulation-density}(c) we also show the band diagram for an InSb shell thickness of 30\,nm in the $n$-type conductivity regime. Due to the wider shell, the energy eigenvalues are formed at lower values compared to the case of $t_\mathrm{s}=22$\,nm depicted in figure~\ref{fig:simulation-density}(b). As a consequence, $n$-type conductance is found at smaller positive gate voltages. The energy states in the wider shell are also more densely distributed. In addition, not only the first but also the second energy subband is occupied by electrons, which gives rise to a higher electron density (cf. figure~\ref{fig:simulation-density}(a)) and a rather broad electron density distribution within the shell. As shown in figure~\ref{fig:simulation-density}(d), the density distribution also has a weak second  maximum located closer to the nanowire center. An important conclusion one might draw from this knowledge is that in these nanowires tubular-like regions might exist which have a lower or higher electron density owing to the varying shell thickness. 

In order to reveal the correspondence between the simulated and measured values of the conductance, one should find the ratio $\kappa$ between $E_\mathrm{F}-E_\mathrm{c}$ and the gate voltage $V_\mathrm{g}$ at the nanowire surface. We follow the simplest way and compare the interval of the voltages/energies corresponding to the depleted regime of the nanowire, i.e. the regime when the conductance is equal to zero. According to figure~\ref{fig:ambipolar}, the gate voltage range where the depletion occurs is about $10$\,V which corresponds to an energy range of about 0.25\,eV in the simulation and $\kappa=0.025$. 

\subsection{Universal Conductance Fluctuations} \label{sec:UCF} 

\begin{figure}[htb]
	\begin{center}
		\includegraphics[width=0.7\textwidth]{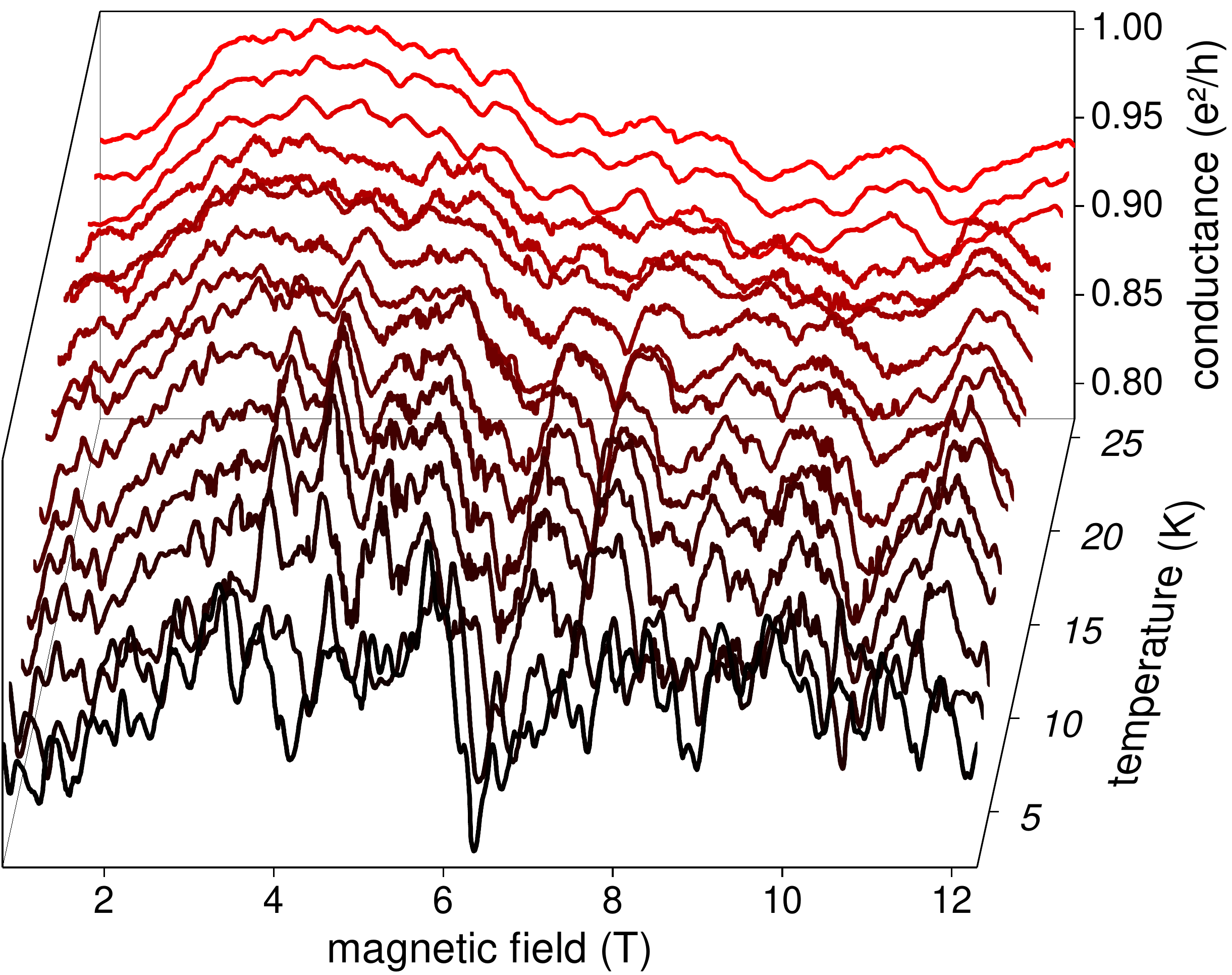}
		\caption{Conductance fluctuations in units of $e^2/h$ of sample B1 as a function of a perpendicular magnetic field at temperatures between 2 and 25\,K.}
		\label{fig:UCF-B-T}
	\end{center}
\end{figure}

After clarifying the general transport behaviour of our core/shell nanowires, we move over to the transport in the presence of a magnetic field. Figure~\,\ref{fig:UCF-B-T} shows a typical conductance measurement of sample B1 as a function of a perpendicular magnetic field at temperatures between 2 and 25\,K. In order to bring the nanowire into the $n$-type regime a back-gate voltage of 30\,V was applied.  
The observed fluctuating pattern of the magnetoconductance can be assigned to universal conductance fluctuations, which originate from a superposition of not-too-many phase-coherent scattering loops of electron waves in the sample \cite{Lee85,Altshuler85c,Lee87,Beenakker88a}. The rich composition of both high and low frequency contributions is an indication for the coexistence of coherent loops with various sizes. If the temperature is increased, the rapidly oscillating components, which are related to larger loops, start to vanish. For temperatures above 17\,K there are no pronounced fluctuations visible anymore, leaving the sample in a state in which the phase-coherence is lost.  

Besides a qualitative description of the spectrum, it is possible to directly 
gain information on the characteristic parameters like the phase-coherence length $l_{\varphi}$. The phase-coherence length $l_{\varphi}$ at different temperatures can be calculated by using the expression \cite{Lee85,Lee87,Beenakker88a}
\begin{equation}
    l_{\varphi}=\gamma\frac{\Phi_0}{2r_\mathrm{t}B_c} \; , \label{eq:lphi}
\end{equation}
with, $\Phi_0$ the magnetic flux quantum $h/e$, $\gamma$ a sample specific pre-factor, $r_\mathrm{t}=r_c + t_\mathrm{s}$ the total nanowire radius, and $B_c$ the correlation field. The latter one is obtained by calculating the autocorrelation function of each single UCF trace. The correlation field acts as a figure of merit for the maximum area $A$ encircled by a closed loop in which the electron partial waves interfere coherently. Following the relation $\Phi_0=B_c\cdot A$, a larger correlation field is related to a smaller loop size and vice versa. Due to the fact that this model was initially developed for the characterization of two-dimensional samples, additional geometric and thermal corrections have to be considered, which are introduced via the pre-factor $\gamma=0.42 \dots 1.3$ \cite{Beenakker88a,Bloemers11}. 

\begin{figure}[htb]
	\begin{center}
		\includegraphics[width=0.8\textwidth]{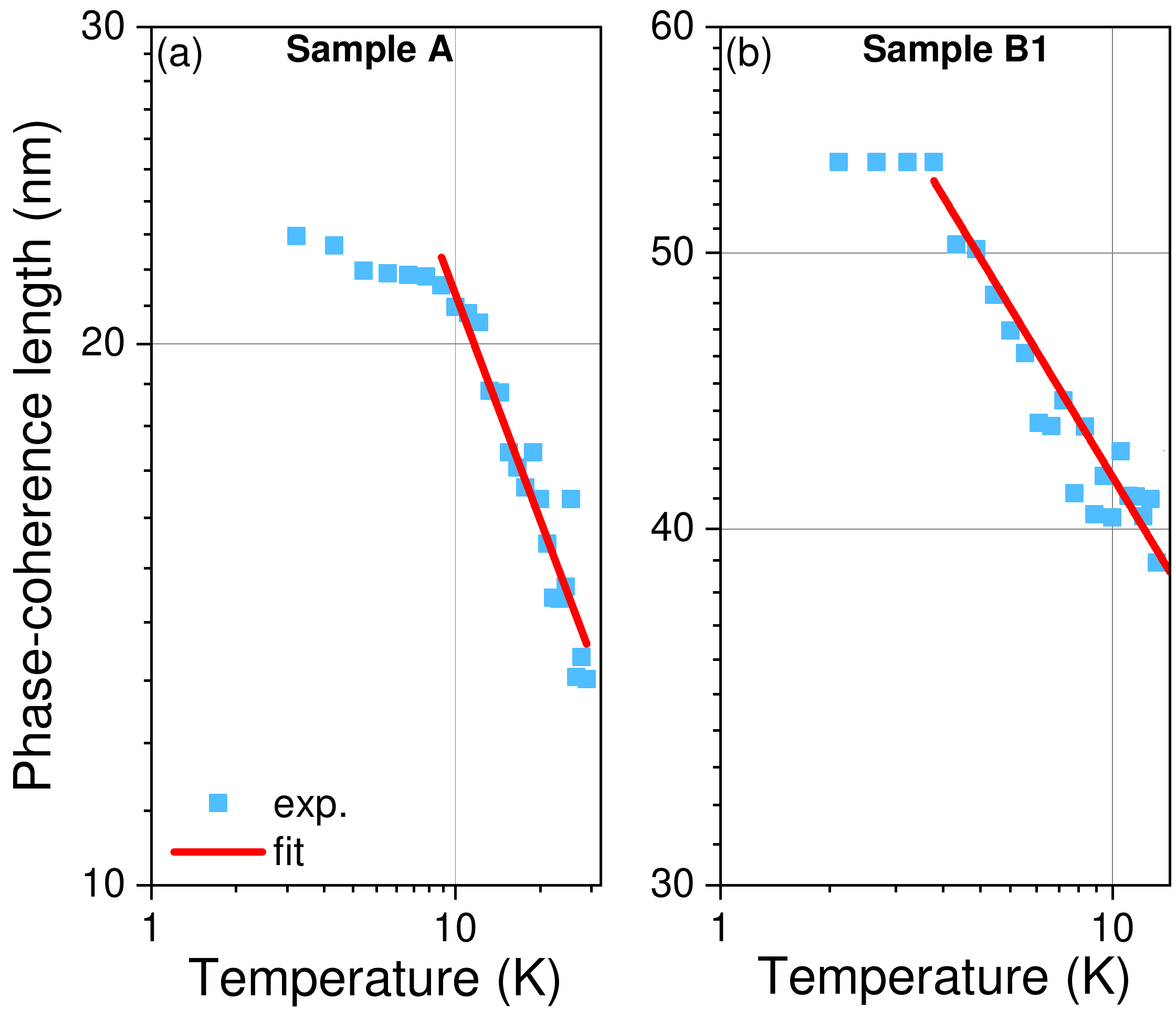}
		\caption{Temperature-dependent phase-coherence length for sample A (a) and B1 (b), based on the analysis of the correlation field B$_c$. The linear fit function (red) is used to extract $\epsilon$ as the damping rate of $l_{\varphi}$(T).}
		\label{fig:UCF-Lphi-T}
	\end{center}
\end{figure}

In figure~\ref{fig:UCF-Lphi-T} $l_\varphi$, determined according to equation~(\ref{eq:lphi}), is shown as a function of temperature for samples A and B1.  
For calculating $l_\varphi$ we have chosen the minimum value for $\gamma=0.42$, which means that the obtained values for $l_{\varphi}$ represent the lower limits of the phase-coherence length. 

Table~\ref{tab:UCF-lphi-overview} gives an overview about the most import parameters regarding the phase coherent transport for sample A and B1. Here, $l_{\varphi ,max}$ refers to the maximum value of the phase-coherence length obtained from the analysis of $B_c$. We attribute the comparably small phase-coherence length to a mismatch between the measured devices and the used model, which assumes a one-dimensional system. In our case, in which the diameter of the sample is in the same range as the contact separation, the exact sample geometry becomes more important \cite{Bloemers11}. Therefore, the pre-factor $\gamma$ and thus $l_{\varphi}$ might differ. However, due to the limited number of datasets, we restrict our further analysis to a qualitative comparison of the individual samples A and B1. The latter one shows a much larger phase-coherence length, which is in good agreement to the, in comparison, outstanding conductance and gate tunability and could be an indication for a better crystal quality and less defects. We would like to stress, that the calculated values for the phase-coherence length just hold for transport along the nanowire axis, i.e. for an out-of-plane magnetic field. Previous measurements on GaAs/InAs core/shell nanowires already revealed, that the phase-coherence length for transport around the nanowire axis, i.e. in-plane fields, is typically 1-2 orders of magnitude larger \cite{Haas16,Haas17a}.

Another important parameter is the temperature-dependent damping $\epsilon$ of the phase-coherence length. Our values of $-0.45$ and $-0.23$, respectively, are thereby in good agreement with previous works on nanowires \cite{Bloemers11,Estevez10}.
\begin{table}[htb]
\begin{center}
\begin{tabular}{|l|l|l|l|l|}
\hline
sample & $r_\mathrm{t}$ (nm) & $L$ (nm) & $l_{\varphi,\mathrm{max}}$ (nm) & $\epsilon$ \\ \hline
A  & 75 & 500 & 23 &    -0.45\\ \hline
B1   & 69      & 500    & 55 & -0.23    \\ \hline
\end{tabular}
\label{tab:UCF-lphi-overview}
\caption{Comparison of the two investigated devices with respect to the sample geometry and characteristic parameters: Total radius $r_\mathrm{t}$, contact separation $L$, and  $l_{\varphi,\mathrm{max}}$ as the maximum phase-coherence length $l_\varphi$ extracted from the correlation field. The fit parameter $\epsilon$ corresponds to the temperature-dependent damping rate of $l_\varphi$.}
\end{center}
\end{table}

\subsection{Weak Antilocalization} \label{sec:WAL}

\begin{figure}[htb]
	\begin{center}
		\includegraphics[width=0.8\textwidth]{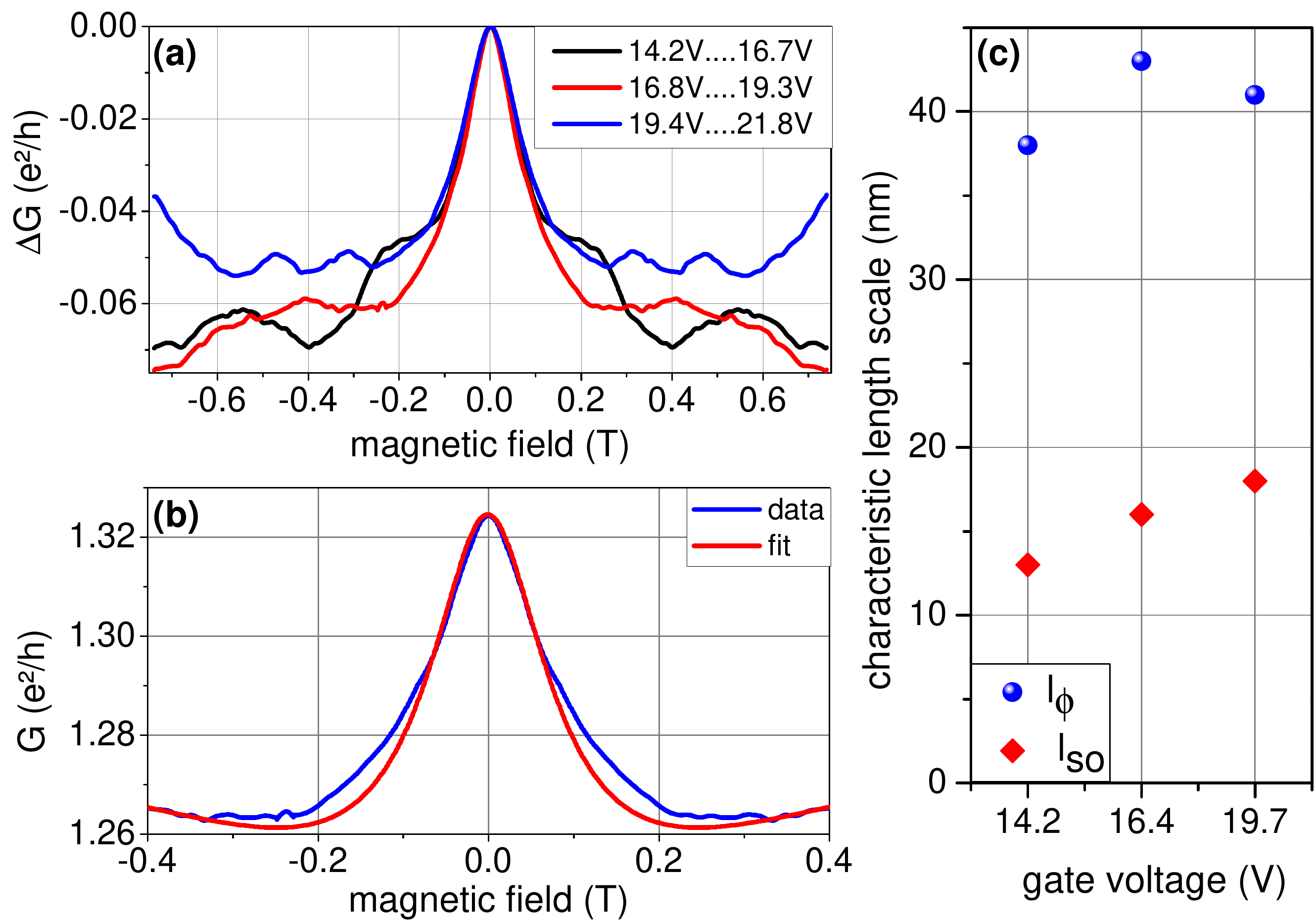}
		\caption{(a) Weak antilocalization features of sample B1 after the gate averaging procedure was performed. Every trace is a combination of 25 individual magnetic field sweeps at different gate voltages in the range of 14.2 up to 21.8\,V. The sections were chosen in a way that the gate-induced increase of the background conductance was much smaller than all corrections related to interference effects. Additionally, in order to  obtain a better comparability, the maximum conductance value of every WAL peak was subtracted from its corresponding conductance trace, i.e. $\Delta G=G(0)-G(B)$ (b) Zoom-in of the second WAL trace, i.e. for $16.8-19.3\,$V, and the resulting fit based on equation~(\ref{eq:DeltaG-WAL}) for $l_{\varphi}=43\,$nm and $l_\mathrm{SO}=16\,$nm. (c) Phase-coherence length $l_\varphi$ and spin-orbit scattering length $l_\mathrm{SO}$ obtained from the fits according to equation~(\ref{eq:DeltaG-WAL}) for the three different gate voltage sections.}.
 \label{fig:WAL}
	\end{center}
\end{figure}

Apart from universal conductance fluctuations, weak localization phenomena can also provide information about phase-coherence  \cite{Bergmann84,Beenakker91c}. Here, depending on weather spin-orbit coupling is involved or not, a magnetoconductance peak or dip is expected at zero magnetic field. Figure \ref{fig:WAL} (a) gives an overview of the conductance of sample B1 for low magnetic fields after the gate averaging procedure is performed for a voltage range of 14.2$\,$V up to 21.8\,V (see Supplementary Material) \cite{Estevez10,Roulleau10}. The averaging was necessary to suppress conductance fluctuations. 
In order to avoid any influence of the change in electron concentration, which is directly translated into a change of the background conductance, the whole gate set is cut into three segments which consist out of 25 individual magnetic field sweeps at different gate positions: $14.2-16.7\,$V, $16.8-19.3\,$V and $19.4-21.8\,$V. All three traces show a clear peak in conductance at $B=0$ indicating weak antilocalization due to the presence of spin-orbit coupling \cite{Hikami80,Bergmann84}. This is in contrast to previously measured GaAs/InAs core/shell nanowires, where a minimum in conductance around zero magnetic field, i.e. weak localization, was observed \cite{Guel14,Haas17a}.

Weak antilocalization and localization, as electron interference effects, both require a sufficiently large phase-coherence length relative to the system geometry. By sharing the same origin, the difference arises due to the significance of the electron spin, expressed by the spin-orbit scattering length $l_\mathrm{SO}$. For a fit of the experimental data to a theoretical model we assume that our wires are in the quasi-one dimensional limit with L$\,\gg\,l_{\varphi}\,\gg \,w$ \cite{Kurdak92,Roulleau10}. For an average carrier concentration and mobility of $n=3\times$10$^{18}\,$cm$^{-3}$ and $\mu=100\,$cm$^2$/Vs, which have been obtained by means of transistor and transconductance measurements for wires of this specific growth run at room temperature \cite{Rieger17}, we get an elastic scattering length of $l_e=v_\mathrm{F} \tau_e=10\,$nm, which sets our system into the dirty metal regime ($l_\mathrm{e} \ll w$) \cite{Beenakker88a}. Thus we can use \cite{Kurdak92} 
\begin{equation}
    \Delta G(B) = -\frac{2e^2}{hL}\left[\frac{3}{2}\left(\frac{1}{l^2_{\varphi}}+\frac{4}{3 l^2_\mathrm{SO}}+\frac{1}{\mathcal{D}\tau_{B}}\right)^{-1/2}-\frac{1}{2}\left(\frac{1}{l^2_{\varphi}}+\frac{1}{\mathcal{D}\tau_{B}}\right)^{-1/2}\right]
    \label{eq:DeltaG-WAL}
\end{equation}
for the conductance correction, with $\mathcal{D}$ the diffusion constant and $\tau_{B}$ the magnetic relaxation time. For the latter one, we take
\begin{equation}
    \tau_{B}=\frac{l^4_{B}}{K_1w^3v_F}+\frac{l^2_{B}\tau_e}{K_2w^2} \; , 
\end{equation}
with $l_{B}\,$=$\,\sqrt{\hbar/eB}$ the magnetic length. Under the assumption of diffusive scattering and by including flux-cancellation effects, we take K$_1\,$=$\frac{1}{4\pi}$ and K$_2\,$=$\frac{1}{3}$ \cite{Beenakker88a,Kurdak92}. Additionally, due to the special wire geometry, which restricts the transport to the conductive InSb shell and forbids all electron trajectories across the nanowire diameter, we make the assumption that $w=45\,$nm, which is equal to the width of the top facet.

Figure \ref{fig:WAL}(b) exemplarly shows a zoom-in of the second WAL trace, i.e. for $16.8-19.3\,$V, and the resulting fit based on equation~(\ref{eq:DeltaG-WAL}), giving $l_{\varphi}=43\,$nm and $l_\mathrm{SO}=16\,$nm. The length of $l_\mathrm{SO}$ is rather small. We mainly attribute this to the small diffusion constant. In figure~\ref{fig:WAL}(c) the phase-coherence length $l_\varphi$ and spin-orbit scattering length $l_\mathrm{SO}$ extracted from the fit are given for the three different gate voltage sections. While $l_\varphi$ is basically constant, we find a small but clear increase of $l_\mathrm{SO}$ with increasing gate voltage. We mainly attribute the presence of spin-orbit coupling to the Rashba effect \cite{Bychkov84}, even though the Dresselhaus contribution in InSb can be rather large \cite{Winkler03}. Indeed, it was shown that, in spite of the fact that the nanowires are grown along the [111] direction, some Dresselhaus contribution might be relevant \cite{Bringer19}. For weak antilocalization in InAs nanowires Rashba as well as Dresselhaus contributions were considered \cite{Kammermeier16,Kammermeier17}. The phase-coherence length extracted here is comparable to the value obtained from the universal conductance fluctuations. 

\subsection{Flux-periodic oscillations} \label{sec:AB}

\begin{figure}[htb]
	\begin{center}
		\includegraphics[width=0.99\textwidth]{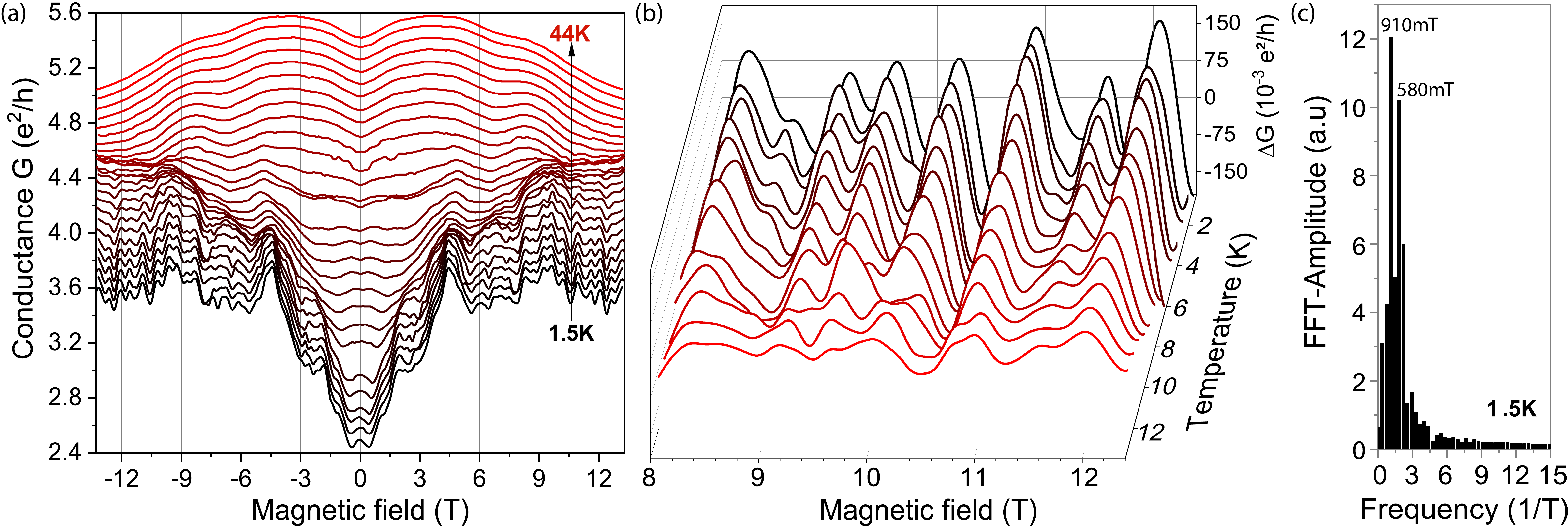}
		\caption{(a) Temperature-dependent magneto-conductance traces for sample~A as a function of an in-plane magnetic field $B_{||}$.(b) Extracted conductance oscillations $\Delta G$ in units of $e^2/h$ for sample~A. The temperature range was restricted to 1.5\,K up to 12\,K. The slowly varying background conductance was subtracted.(c) Fast-Fourier-Transform (FFT) spectrum of the flux-periodic oscillations at T\,=\,1.5\,K.}
		\label{fig:AB-PZ08-A5-waterfall}
	\end{center}
\end{figure}

In addition to measurements in a perpendicular magnetic field we also performed transport measurements in a magnetic field aligned to the nanowire axis.Figure~\ref{fig:AB-PZ08-A5-waterfall}(a) shows the magneto-conductance of sample A at temperatures ranging from 1.5 to 44\,K. Due to symmetry reasons, i.e. $G(B)=G(-B)$, we can conclude that the Onsager reciprocity is fulfilled \cite{Büttiker86}. It can be clearly seen, that all traces up to approx. 25\,K are modulated by slowly varying background fluctuations, which can be attributed to universal conductance fluctuations. The latter ones originate from scattering loops with small areas, which are cooped within the InSb section of the wire and therefore limited by the thickness of the shell. However, for temperatures up to 12\,K, the slowly fluctuating background is superimposed by flux-periodic oscillations. In order to analyze these regular oscillations, the low-frequency contribution of the universal conductance fluctuations was subtracted by means of a Savitzky-Golay fitting routine (see Supplementary Material). Furthermore, we only considered a range at larger magnetic fields in order to skip the effect of weak antilocalization. The resulting magneto-conductance traces, depicted in Figure~\ref{fig:AB-PZ08-A5-waterfall}(b), show a clear Aharonov--Bohm type oscillatory behaviour. Upon increasing the temperature, the oscillation amplitude is decreasing. The oscillations can be attributed to flux-periodic oscillations, as previously observed in nanowires with a conductive InAs shell around a GaAs core \cite{Guel14,Haas16,Haas17,Haas17a}. These oscillations are attributed to closed-loop quantized states located within the InSb shell which shift their energetic position periodically with the encircled magnetic flux. The oscillation period is determined by the magnetic flux quantum and the cross section of the loop
\begin{equation}
\Delta B =\Phi_0/(\pi r^2) \; . \label{eq:DeltaB}
\end{equation} 
Here, we assumed a circular area which is encircled by the phase-coherent loop state. The Fourier transform of the magneto-conductance reveals peaks at $1.1\,\mathrm{T}^{-1}$ and $1.6\,\mathrm{T}^{-1}$, as shown in Figure~\ref{fig:AB-PZ08-A5-waterfall}(c), which correspond to a radius of 38 and 45\,nm, respectively. Since sample A has a core radius of 25\,nm and a shell thickness of 50\,nm, we can conclude, that the closed-loop states are located within the InSb shell, but slightly shifted towards the center. The presence of two peaks in the Fourier transform, i.e. two different cross sectional areas, can have two reasons. First, there might be some potential fluctuations in the shell due to defects or, second, the large $g$-factor of InSb might lead to an additional splitting of the energy spectrum \cite{Rosdahl14}.      

\begin{figure}[htb]
	\begin{center}
	\includegraphics[width=0.9\textwidth]{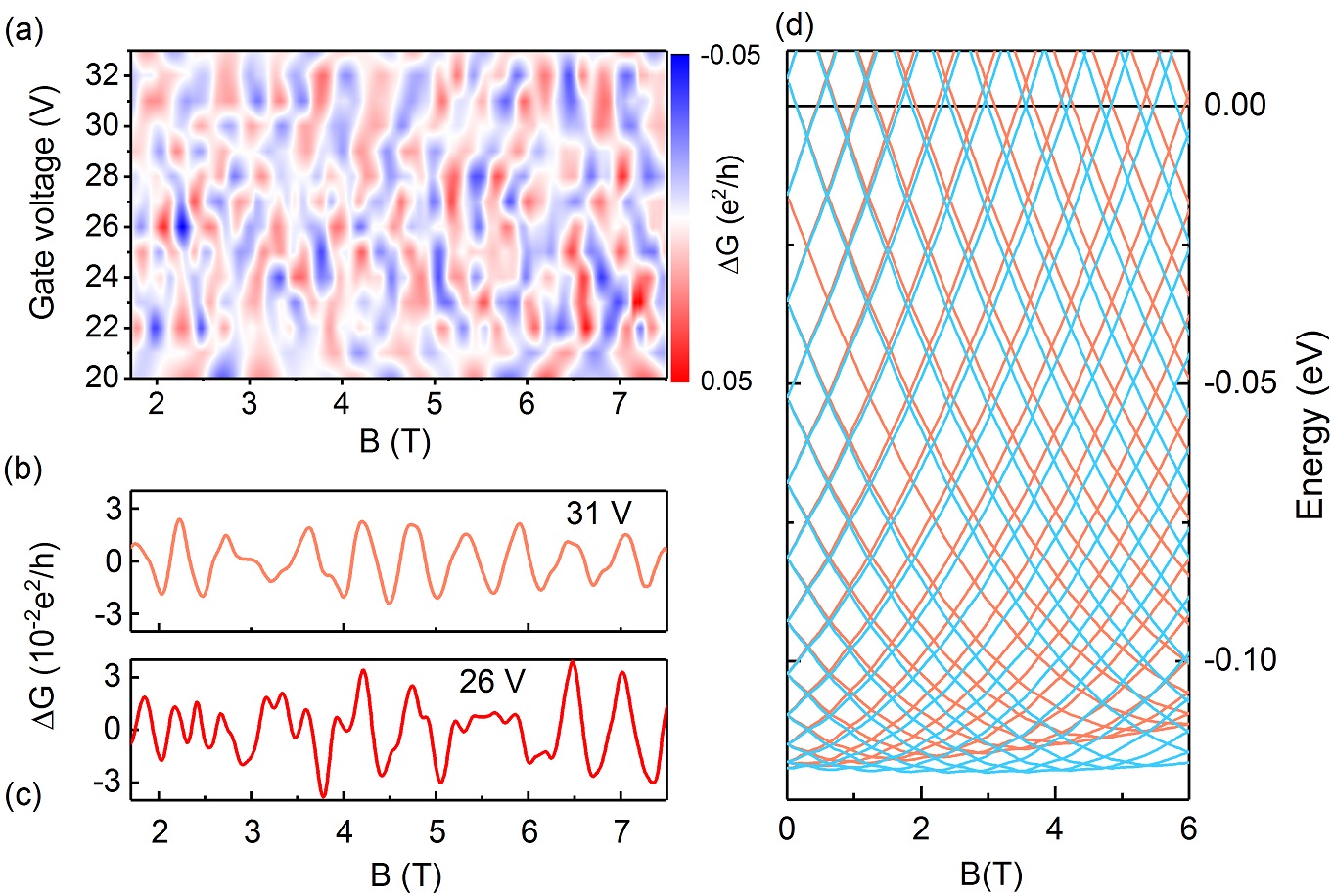}
		\caption{(a) Two-dimensional color-scale plot of the oscillating fraction of the magneto-conductance $\Delta G$ vs. the applied gate voltage and the parallel magnetic field for sample B2. The slowly varying background conductance was subtracted from the total conductance. (b), (c) Measured $\Delta G (B)$ traces as a function of $B$ for $V_\mathrm{g}=31$\,V and $26$\,V, respectively. (d) Calculated energy levels for spin up and down states for a nanowire corresponding to growth run B with $r_c=35$\,nm, $t_\mathrm{s}=22$\,nm, and $(E_\mathrm{F}-E_\mathrm{c})=0.225$\,eV.}
		\label{fig:AB-gate-exp-simu}
	\end{center}
\end{figure}

In order to get a deeper insight into the origin of a certain oscillation pattern we performed magneto-conductance measurements in an axial magnetic field as a function of back-gate voltage. In figure~\ref{fig:AB-gate-exp-simu}(a) corresponding measurements of $\Delta G$(B) are shown for sample B2. 
Once again one finds regular conductance oscillations as a function of the magnetic field. However, upon varying the gate voltage, the phase of the oscillations and also the oscillation pattern changes. In order to see this more clearly single traces of $\Delta G (B)$ are shown in figures~\ref{fig:AB-gate-exp-simu}(b) and (c) for a gate voltage of 31\,V and 26\,V, correspondingly. Obviously, in the first case the magneto-conductance oscillates more slowly and regularly than in the second case. 

The measured magneto-conductance oscillations are analyzed in more detail by calculating the energy spectrum as a function of magnetic field. In the same spirit as the analysis presented in Sec.~\ref{sec:ambipolar}, the carrier density at the Fermi level $n_\mathrm{F}(B)$ is assumed to be proportional to the conductance variations $\Delta G(B)$. Figure~\ref{fig:AB-gate-exp-simu}(d) shows the calculated energy spectrum as a function of an axial magnetic field for a nanowire structure corresponding to growth run B, i.e. assuming a shell thickess $t_\mathrm{s}=22$\,nm. The magnetic field lifts the degenerary of the orbital momentum quantum number $l$ and the energy corresponding to each orbital momentum quantum number is almost parabolic with respect to the magnetic field. Thus, without spin effects, the energy spectrum is quasi-periodic with the magnetic field with a period $\Phi_0/(\pi r_{av}^2)$. Here, $r_{av}$ is the average electronic radius which approximately corresponds to the distance from the maximum of the electron density distribution to the nanowire center (cf. figure~\ref{fig:simulation-density}(d)). The effect of the electron spin is modeled by introducing the Zeeman potential into the Schr\"odinger equation using a bulk InSb $g$-factor of $g_\mathrm{InSb}=-51$ \cite{Weperen13}. The Zeeman effect breaks spin degeneracy increasing (decreasing) the energy of the states with spin down (up). The energy spectrum becomes modulated and fairly more complicated. We neglected the effect of spin-orbit coupling.

We assume, that the wire is in the diffusive regime and that only small bias voltages, i.e. $V_\mathrm{sd} < 0.1\,$mV, are applied. Thus, only electrons in close vicinity to $E_\mathrm{F}$ can participate in the transport. In that case the magneto-conductance is proportional to the number of carriers at the Fermi level \begin{equation}
G(B) \sim n_\mathrm{F}(B) = \int_{E_\mathrm{F}-\Delta}^\infty D_\mathrm{1D}(E,B) \frac{dE}{1 + \exp [(E-E_\mathrm{F})/k_\mathrm{B}T]} \; ,    
\end{equation}
where $D_\mathrm{1D}(E,B)$ is the one-dimensional density of states at a given magnetic field $B$, and $k_\mathrm{B}$ is the Boltzmann constant. The assumed energy spreading $\Delta$ is about several $k_\mathrm{B}T$, with $T=4$\,K. As the magnetic field increases, the number of orbital momentum states in the vicinity of the Fermi level varies quasi-periodically, which ultimately leads to the oscillations in $G(B)$. 

\begin{figure}[htb]
	\begin{center}
\includegraphics[width=0.9\textwidth]{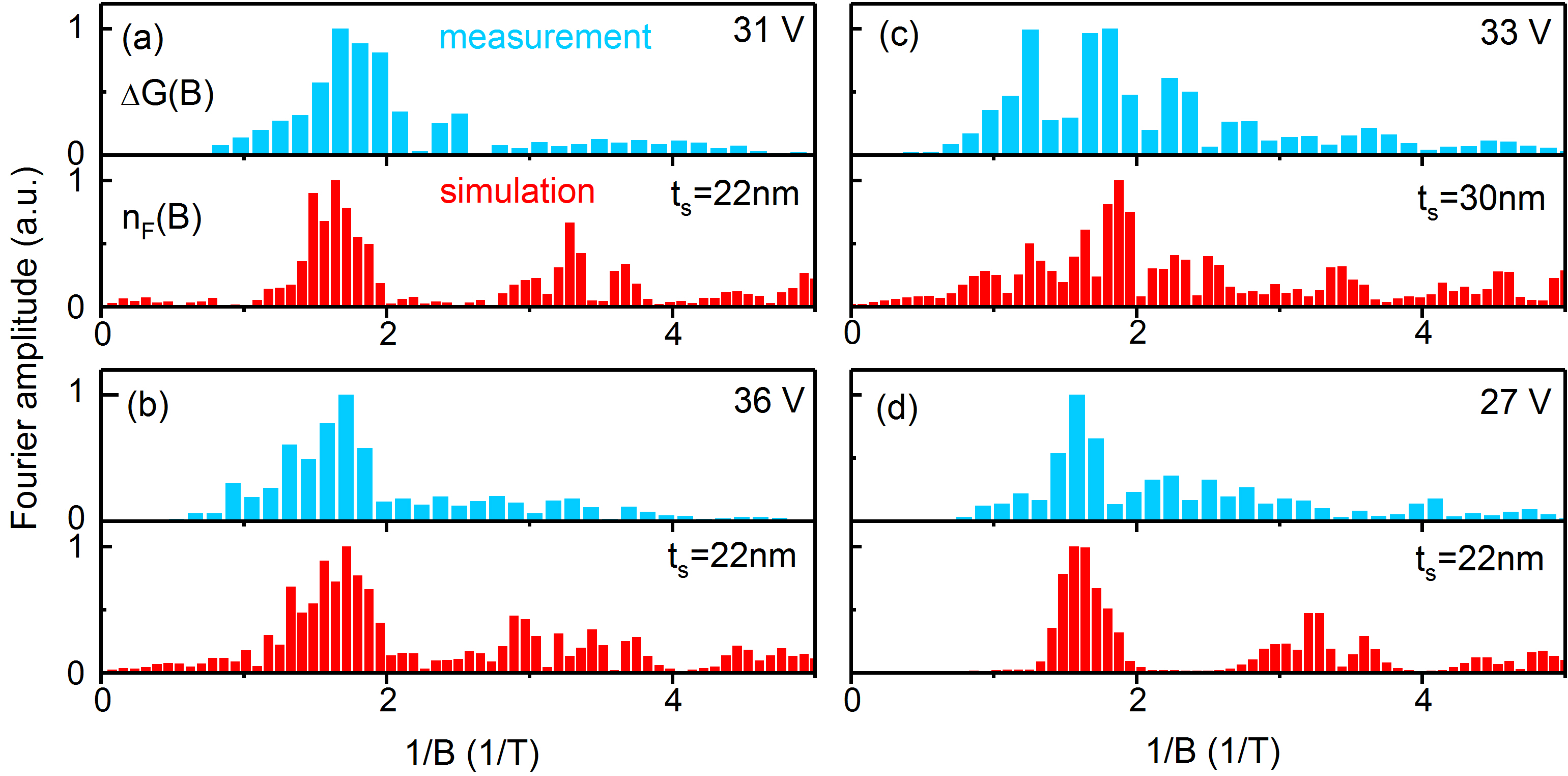}
		\caption{Normalized Fourier transform of the oscillating measured magneto-conductance $\Delta G(B)$ (blue) and the calculated electron density $n_\mathrm{F}(B)$ (red), the gate voltage is set to 31\,V (a), 36\,V (b), 33\,V (c), and 27\,V (d), respectively.}
		\label{fig:simulation-AB-FFT}
	\end{center}
\end{figure}

In order to analyze the frequency components of the magneto-conductance oscillations we performed Fourier transforms of the experimentally obtained traces of $\Delta G(B)$ and of the simulated densities $n_\mathrm{F}(B)$. Some representative spectra are displayed in figure~\ref{fig:simulation-AB-FFT}. 

The frequency analysis of the experimental data reveals three dominant types of spectra. The first type (cf. figure~\ref{fig:simulation-AB-FFT}(a) and (b), upper panels) is the most frequently observed one over the measured gate voltage range and has a peak at a frequency of about $1.7$\,T$^{-1}$. According to equation~(\ref{eq:DeltaB}), we can conclude, that the oscillations are caused by the electron ensemble with average radius of $r_{av} \sim 47$\,nm. Note that the nanowires of growth run B have a core with the radius of about 35\,nm and a shell width varying along the nanowire axis. The calculation of the spatial electron distribution in the nanowire with varying shell thickness shows that $r_{av}$ of 47\,nm assigns to the electron ensemble in the nanowire with the shell thickness of about 22\,nm. The corresponding electron distribution is shown in figure~\ref{fig:simulation-density}(d).  Armed with this knowledge we calculate the oscillations of $n_\mathrm{F}(B)$ with the magnetic field for a shell thickness of 22\,nm and performed the corresponding Fourier transform. They are presented in figure~\ref{fig:simulation-AB-FFT}(a) and (b), lower panels, and agree well with the experimental data. We would like to stress that, in accordance with the calculations for the higher gate voltage of 36\,V, already electrons of not only the first ($n=1$), but also the second subband ($n=2$) contribute to the magneto-conductance. However, the frequency of the peak in the Fourier transforms of $\Delta G(B)$ and $n_\mathrm{F}(B)$ (cf. figure~\ref{fig:simulation-AB-FFT}(b)) remains at about $1.7$\,T$^{-1}$, i.e. it does not change with the gate voltage. The second type of the spectra taken form the experimental data shown in figure~\ref{fig:simulation-AB-FFT}(c) takes up a much broader frequency range than the first type. Basically it has no very distinct sharp peak. It rather shows some distribution of frequencies which might be attributed to the electron ensemble with a wider space distribution than in the case of the stronger confinement we discussed above. This kind of distribution is typical for the electron ensemble in a nanowire with the wider shell, i.e. for a shell thickness of 30\,nm, as shown in figure~\ref{fig:simulation-density}(d). The calculated Fourier transform of $n_\mathrm{F} (B)$ for this shell thickness is displayed in figure~\ref{fig:simulation-AB-FFT}(c) and has certain similarities with the experimental data. The third type of the spectra (cf. figure~\ref{fig:simulation-AB-FFT}(d)) has a distinct peak at the frequency of about $1.7$\,T$^{-1}$ and also some frequency components forming a long tail. We attribute the pronounced peak to an electron ensemble confined within the narrow shell with the width of 22\,nm. The corresponding calculated Fourier transform of $n_\mathrm{F}(B)$ is shown in figure~\ref{fig:simulation-AB-FFT}(d). The position of the peak fits to the experimental data. The additional frequency components might be related to electrons in shell sections having a bigger radius.

The analysis of the Fourier spectra of the oscillating magneto-conductance suggest that, depending on the applied gate voltage, the oscillations are provided by electron fractions located in different regions of the nanowire. We may expect that due to the fact that the phase-coherence length ($l_\varphi \approx 50$\,nm) is smaller than the distance between the drain and source contacts (350\,nm), i.e. the coherent orbital momentum states relevant for a particular oscillation pattern are formed in a region of the nanowire shifting along the nanowire axis with the gate voltage. Probably, this might occur also during the measurements of the magneto-conductance in GaAs/InAs core/shell nanowires performed during the previous study \cite{Guel14,Haas17a}. However, as the shell thickness has been uniform along the nanowire axis, in this case there was no clear evidence of this effect while now we may confirm this. We did not find a clear evidence of Zeeman splitting in our experiments, although this was observed in the simulations at some specific gate voltages by showing a splitted peak in the Fourier spectrum. We suspect, that the impact of the Zeeman effect is masked by other effects, e.g. varying shell thicknesses which results in a broadening of the Fourier spectrum.   

\section{Conclusions} \label{sec:conclusion}

In conclusion, we could show that in nanowires with a low-band gap InSb shell wrapped around a large band gap GaAs core a highly conductive channel in the shell can be created. In contrast to an InAs shell the carrier transport could be altered from $n$-type to $p$-type by means of a back gate. We attributed this special feature to the low band gap energy of InSb. At low temperatures, electron interference effects, i.e. universal conductance oscillations were observed in an external magnetic field perpendicular to the nanowire axis. Using these measurements, we could estimate the phase-coherence length $l_\varphi$. Especially the presence of weak antilocalization shows that spin-orbit coupling plays a significant role in the InSb shell. Moreover, the core/shell geometry also allowed us to observe Aharonov--Bohm-type oscillations in magnetic fields along the nanowire axis, which are periodic with the magnetic flux quantum. This directly proves that phase-coherent closed-loop states are formed in the InSb shell. The Aharonov--Bohm type oscillations provide an additional control knob to adjust the nanowire conductance in a well-controlled manner in addition to applying a gate voltage. Flux control is particularly interesting when the nanowire is combined with superconducting electrodes \cite{Guel14}, e.g. in connection with Majorana physics. Although the $g$-factor of InSb is expected to be exceptionally large, we did not find a clear signature of Zeeman splitting in the Aharonov--Bohm type oscillations, yet. We attribute this to variations in the shell diameter. This leads to an addition modulation of the oscillations which makes it difficult to assign deviations from regular flux-periodic oscillations unambiguously. Thus, for future studies it might be interesting to further optimize the growth of the shell to obtain a shell with a more homogeneous shell thickness along the wire axis. This would reduce the according frequency modulations and by that enhancing the chance to single out spin-related effects.  

\section{Acknowledgement}

All samples were prepared at the Helmholtz Nano Facility \cite{GmbH2017}. We thank H. Kertz for assistance during the measurements. 

\newpage
\bibliographystyle{unsrt}
\bibliography{main.tex}
\end{document}